\documentclass{acm_proc_article-sp}

\usepackage{amssymb,amsfonts,amsmath,color}
\usepackage{bbm}

\newcommand{\superscript}[1]{\ensuremath{^{\textrm{#1}}}}
\def\sharedaffiliation{\end{tabular}\newline\begin{tabular}{c}}

\def\rbi{\superscript{1}}
\def\eth{\superscript{2}}
\def\fer{\superscript{3}}
\def\sig{\superscript{4}}

\begin{document}

\title{Modeling peer and external influence in online social networks}

\numberofauthors{5} %  in this sample file, there are a *total*
\author{
  Matija Pi\v{s}korec\rbi, Nino Antulov-Fantulin\rbi\superscript{,}\eth, Iva Miholi\'{c}\fer, Tomislav \v{S}muc\rbi, Mile \v{S}iki\'{c}\fer\superscript{,}\sig\titlenote{corresponding author} \\
  \sharedaffiliation
  \begin{tabular}{c}
    \affaddr{{\rbi}Laboratory for Machine Learning and Knowledge Representations, Ru\dj er Bo\v{s}kovi\'{c} Institute, Zagreb, Croatia} \\
    \affaddr{{\eth}Computational Social Science, ETH Zurich, Switzerland} \\
    \affaddr{{\fer}Faculty of Electrical Engineering and Computing, University of Zagreb, Zagreb, Croatia} \\
    \affaddr{{\sig}Bioinformatics Institute, A*STAR, Singapore, Republic of Singapore} \\
  \end{tabular}
  \sharedaffiliation
  \begin{tabular}{c}
    \affaddr{ \{matija.piskorec, nino.antulov.fantulin, tomislav.smuc\}@irb.hr, \{iva.miholic, mile.sikic\}@fer.hr } \\
  \end{tabular}
}

\maketitle

\begin{abstract}

Opinion polls mediated through a social network can give us, in addition to usual demographics data like age, gender and geographic location, a friendship structure between voters and the temporal dynamics of their activity during the voting process. 
Using a Facebook application we collected friendship relationships, demographics and votes of over ten thousand users on the referendum on the definition of marriage in Croatia held on $1^{st}$ of December 2013.
We also collected data on online news articles mentioning our application.
Publication of these articles align closely with large peaks of voting activity, indicating that these external events have a crucial influence in engaging the voters.
Also, existence of strongly connected friendship communities where majority of users vote during short time period, and the fact that majority of users in general tend to friend users that voted the same suggest that peer influence also has its role in engaging the voters.
As we are not able to track activity of our users at all times, and we do not know their motivations for expressing their votes through our application, the question is whether we can infer peer and external influence using friendship network of users and the times of their voting.
We propose a new method for estimation of magnitude of peer and external influence in friendship network and demonstrate its validity on both simulated and actual data.

\end{abstract}



 % NOT required for Proceedings

\section{Introduction}

Rising popularity of social networks allows us to investigate dynamics of social interactions on a scale that would be unimaginable just a couple of decades ago~\cite{Borge-Holthoefer2011,Kramer2014,Lewis2008,Karsai2014,Guille2013,DeDomenico2013,Najar2012,Yang2010}.
For example, if we are conducting a survey in traditional way we are heavily restricted with the number of participants we are able to reach, and the type of data we can collect.
On the other hand, using a social network as a mediating platform for a survey allows us to gather (in addition to usual demographics data like age, gender and geographic location) a friendship structure between voters and the temporal dynamics of their activity during the voting process.

Using a Facebook application we collected demographics data and votes of $11538$ Facebook users on the referendum on the definition of marriage in Croatia held on $1^{st}$ of December 2013. 
The voters were asked the question: ``Are you in favor of the constitution of the Republic of Croatia being amended with a provision stating that marriage is a community between a woman and a man ?''.
Application was active during a week prior to the referendum and it allowed users to express their voting preference for the upcoming referendum, to see global statistics for all users who voted, and to see statistics for their friends who voted.
In addition, they could also share the link to the application through Facebook.
For all these users we have their friendship relationships and various demographics data like age, gender and geographic location.
Due to the politically charged topic, the referendum attracted a lot of media attention, with the opposing sides trying to engage voters through both classical news media and social media~\cite{Romero2011}.
As we expected that the information about our application would gradually spread throughout both channels, we also collected publication times of major online news articles mentioning our application and the number of visitors coming from these web sites.

\begin{figure}
\centerline{\includegraphics[width=0.5\textwidth]{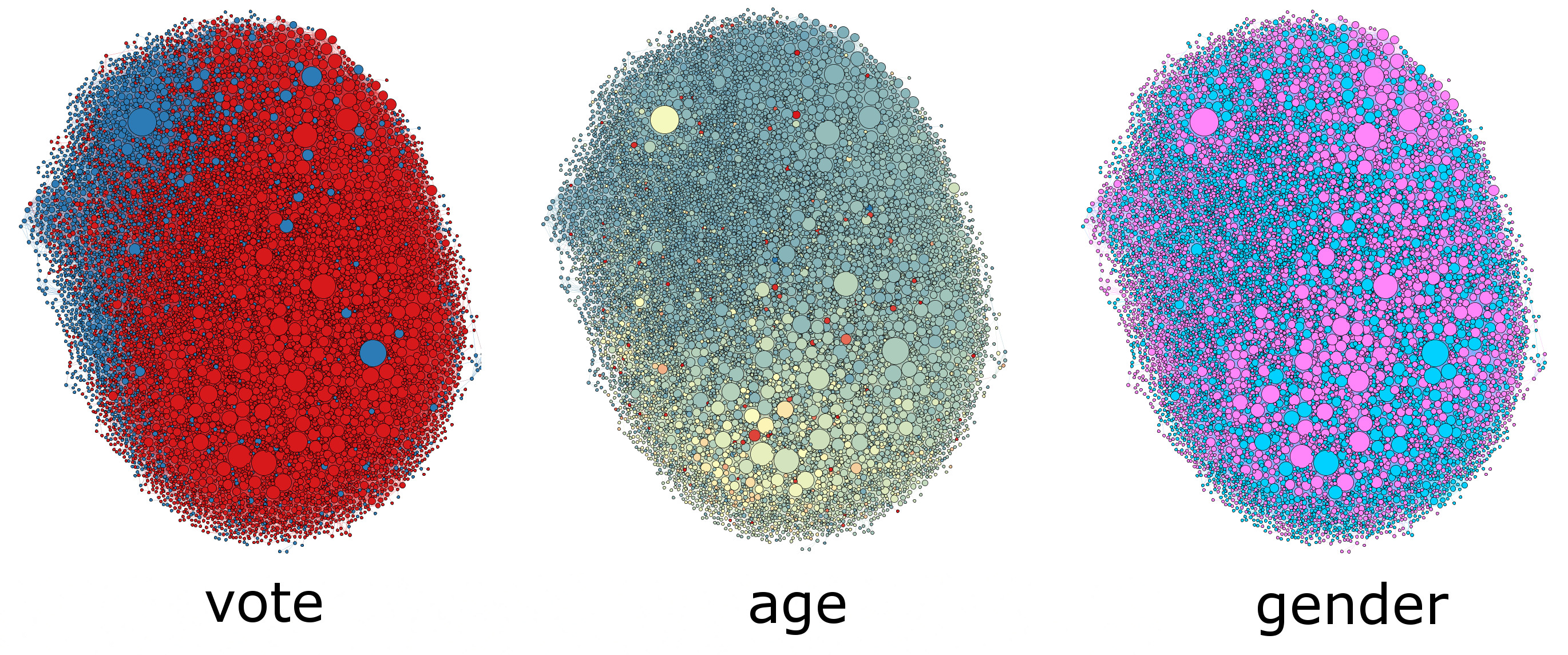}}
\caption{
Network of Facebook users who voted on our application colored by three attributes - vote, age and gender. Votes network are colored blue for ``for'' votes and red for ``against'' votes. Age network is colored pale blue for for young voters (18-30 years of age), pale yellow for middle age voters (30-50 years of age) and orange-red for old voters (over 50 years of age). Gender network is colored pink for female voters and blue for male voters.
}
\label{fig:networks}
\end{figure}

Figure~\ref{fig:networks} shows resulting friendship networks colored by various attributes we collected.
We observe strong homophily regarding the user's votes - users tend to have more friends who voted the same than the ones who voted opposite. 
Some other attributes like age and geographic location also show strong homophily, while others, like gender, show none.
Community analysis on the friendship network reveals that each community is highly homogeneous regarding the votes and that they usually contain couple of highly connected individuals.
We model voting activity dynamics in order to assess whether peer influence or external influence better explains the activation of voters. 
The word ``influence'' here refers to the influence that either peers or some external force like news media play in engaging the users to vote on our application.
We are not interested in the question of how social influence determines the attitudes of individuals, that is, whether users tend to friend each other based on their preexisting preferences or they just become more similar over time~\cite{Lewis2012,Zhang2014}.
Regarding the external influence, we observe that large peaks in voting activity align with the publication times of major online news articles, indicating that media plays a crucial role in engaging the voters. 
On the other hand, some peaks in voting activity are not aligned with any of the publication times of major online news articles. 
For them we observe that majority of votes came from a particular community of highly connected users, indicating that they are mainly driven by the peer influence. 
We propose a methodology that enables us to estimate magnitude of peer and external influence in network using activation cascade.

The main contributions of this paper are the following:  
(i) We collected and described a large temporal Facebook network of social engagement between users.
(ii) Our analysis shows strong homophily with respect to votes in the network, both on the local level (users tend to friend other users who voted the same) and the mesoscale level (communities of friendships are mostly homogeneous with respect to the votes of their users).
(iii) We propose a method for estimation of magnitude of peer and external influence in network by using the activation cascade.
%(iv) Using our method we show that both peer and external influence can play crucial roles in engaging the voters, although the external influence is more dominant.

\section{Related work}
There are decades of research originating from social science on the evolution of social networks~\cite{Snijders2010} and social contagions~\cite{Granovetter,Watts2002}.
Information or rumor spreading were historically modeled via epidemic-like stochastic processes on networks with the Daley-Kendall \cite{DKmodel} and the Maki-Thompson \cite{MTmodel} model, where nodes can be in three states (Ignorant, Spreading, Stifler). 
The simpler stochastic version with the binary state dynamics (active vs. non-active) is the Independent Cascade Model \cite{ICM}, where the active nodes independently try to activate neighboring nodes with certain constant probability.  
The Linear Threshold model \cite{Granovetter, Kempe} models the activation as a weighted sum of active neighbors over a node activation threshold.  
Shen et. al. have made the first attempt of modeling the information propagation \cite{Shen2012} in multi-level networks by Linear Threshold Model in Twitter-Foursquare networks and academic collaboration multiple networks. 

Main problem is how to distinguish true influence in social networks, or what is usually called \emph{social causation}, from correlation effects which derives from homophily or external confounding factors. 
Several things make this ask much easier: (i) if information that is shared is as specific as possible, for example when sharing specific url's instead of generic tags, and (ii) if information that is shared can not be obtained externally, or ideally can be obtained from just few external sources, as otherwise it is very probable that multiple users will somehow acquire the same information independantly, and any potential social influence will be overhelmed with this confounding effect. 
In our case we don't know explicitaly who shared an information with whom and when, so we have to resort to causation vs correlation analysis that we perform by using similar randomization strategy as in~\cite{Anagnostopoulos2008}.

Information can propagate not only over a network (peer propagation) but also via other external channels like mass media.
In fact, large information cascades in social networks are often driven by exogenous events, including political unrest~\cite{Borge-Holthoefer2011,González-Bailón2011} and natural disasters~\cite{Lu2014}.
Peer and external influence can be defined on the level of users, where we are interested to what extent are users influenced by factors internal or external to the network~\cite{Myers2012,Hill2010,Aral2009}, or on the level of items~\cite{Agrawal2012}, where we are interested to what extent is the spread of an item due to factors that are internal or external to the network.
Anagnostopoulos et. al. use a logistic regression to quantify the extent of peer and external pressure on the observed information cascades \cite{Anagnostopoulos2011}. 
Probability of activation can also be modeled with additional introduction of an exposure curve which quantifies relationship between number of exposures coming from friends and the probability of activation~\cite{Myers2012,Myers2012clash,Romero2011}. 
Contrary to the Anagnostopoulos et. al., we take into consideration the decay of influence in time. 
Furthermore, we try to decouple the external and peer influence just by using a statistical properties of activation cascades on network without inferring the actual exposure curves.
Due to the efficiency constraints of analyzing large information cascades, some approaches try to avoid direct calculation on the actual networks by including the network structure implicitly~\cite{Yang2010} or rely on some network statistic like degree distribution~\cite{Brach2014}.

\section{Dataset}

Online social networks provide an opportunity to collect large amounts of data, but due to their nature they provide challenges to experimental design~\cite{Walker2015}.
Usually, a researcher needs to make a tradeoff between conducting an observational study without explicit consent from the users, which raises ethical concerns~\cite{Kramer2014,Verma2014}, or conducting a study where explicit consent is mandatory, which restricts the amount of data that can be collected.
Even when researchers have a direct access to the whole social network and are in position to present their experiment automatically to the large number of users it is still not straightforward to collect large number of responses. 
For example, a study from Aral and Walker~\cite{Aral2012} on a sample of 1.3 million Facebook users managed to collect responses of only 7730 users. 

We used a Facebook application as an online poll for the upcoming referendum, for which an explicit consent for participation in the study had to be given by each Facebook user.
After they expressed their votes, users could see global statistics for all users who voted, and see statistics for their friends who voted.
The full dataset consists of the friendship network of $12695$ Facebook users who registered on our application, along with their age, gender and geographic location (locality).
Out of these, $11538$ Facebook users voted through our application.
From these we consider only users that voted since the opening of the application at $1:24$ AM, $25^{th}$ of November $2013$, until the end of the day of the actual referendum at $11:59$ PM, $1^{st}$ of December $2013$.
Additionally, we extract giant component of their friendship network to obtain $10175$ users on who we make most of the analysis presented in this paper. 

% We also collected full friends lists of all registered users, totaling in $1695421$ unique Facebook usernames in our dataset.
% Because some unregistered users appear in multiple friends list we are also able to retrieve a small part of their local friendship relations.
%As we consider this to be a very valuable dataset about social engagement of online users mediated through social network we decided to make part of this dataset available on GitHub\footnote{\texttt{http://lis.irb.hr/referendum2013/}}.
%The source code of the Facebook application through which data was collected is also available on GitHub\footnote{\texttt{https://github.com/devArena/referendum2013.hr}}

\subsection{Homophily in network}
\begin{figure}
\centerline{\includegraphics[width=0.5\textwidth]{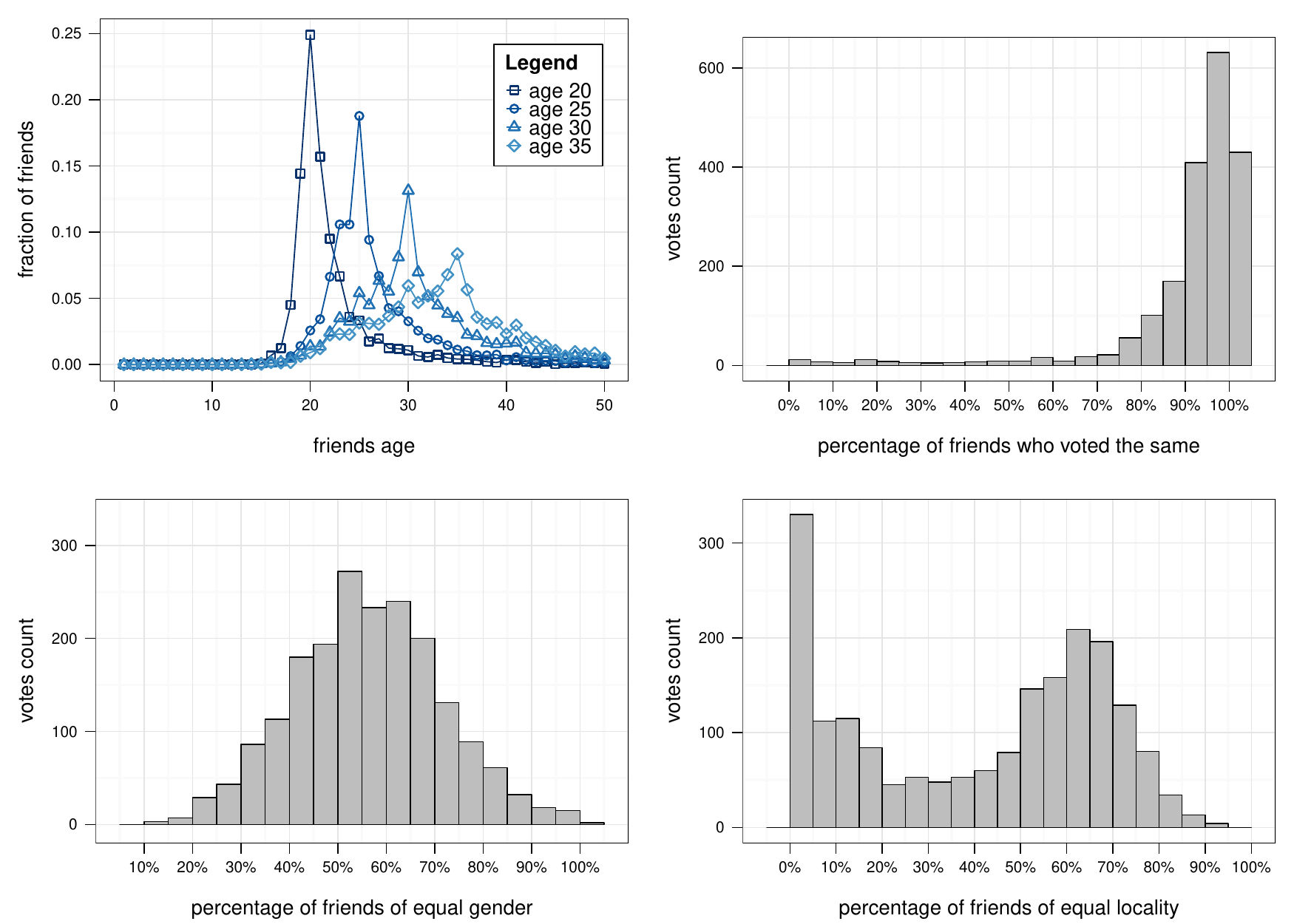}}
\caption{
Homophily in network.
Top left panel shows homophily with respect to age - users are more likely to friend users that are closer their age.
Top right panel shows homophily with respect to votes - users are much more likely to friend users with equal voting preference. 
Similar, bottom right panel shows homophily with respect to geographic location - users are more likely to friend users that are in the same location. 
In contrast, bottom left panel shows that homophily with respect to gender is not so strongly expressed - users are equally likely to friend people of both genders.
}
\label{fig:homophily}
\end{figure}

Simple exploratory analysis of network of voters immediately reveals large homophily with respect to votes, location and age, as seen on Figure~\ref{fig:homophily}. 
Homophily with respect to votes is the strongest, with majority of users having $80\%$ or more friends who voted the same as they did.
This gives us confidence that there is a strong peer-mediated influence that is crucial in spreading the information on our application.
Later we confirm this by analyzing community structures in network and their voting dynamics.
Homophily with respect to age is also strong, especially for younger users. 
This observation is consistent with study performed on much larger Facebook network~\cite{Ugander2011}.
In comparison, homophily with respect to gender is not present, with users being equally likely to friend users of both gender.

\subsection{Communities of voters -- peer influence}
\label{sec:communities}
\begin{figure}
\centerline{\includegraphics[width=0.5\textwidth]{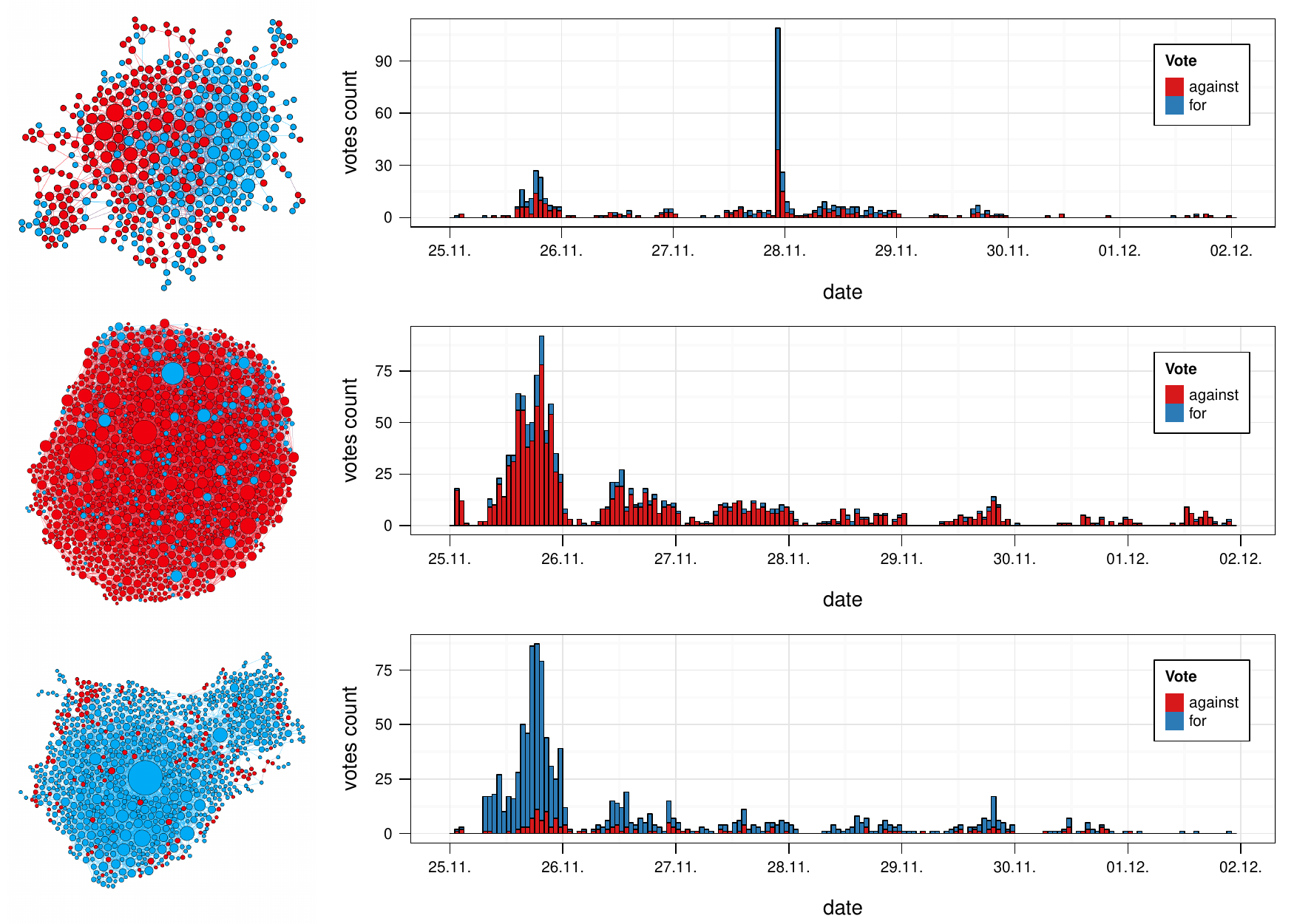}}
\caption{
Friendship communities between people who voted obtained with multilevel algorithm for community finding~\cite{Blondel2008}, colored by their votes (red for ``against'' and blue for ``for'' votes).
Panels on the right show hourly activity throughout the week. Bottom two communities are typical in respect that they are highly homogeneous with respect to their votes, and that they have couple of highly connected users. 
Community in the top panel is an interesting exception because it has almost equal number of votes for each side, and has no highly connected users. 
This community also exhibits interesting voting dynamic because majority of its users voted during one particular hour on the evening of $27^{th}$ of November.
Our analysis shows that this peak in activity is characteristic for this community only, which makes it highly likely that it originated because of the peer-driven influence inside this community.
We consider this to be a purely \emph{peer} driven effect and as such we use it as a golden standard in detecting peer influence.
}
\label{fig:community}
\end{figure}

Using multilevel algorithm for community finding~\cite{Blondel2008} in the software package igraph~\cite{Csardi2006} we detected $27$ communities in our network.
As suggested by the strong homophily with respect to the votes, we found that majority of these communities are also very homogeneous with respect to the votes.
Also, their voting dynamics are all very similar: they reflect global voting dynamics, and they usually contain few strongly connected users, as seen on Figure~\ref{fig:community}.
Notable exception is a community that has almost equal number of votes for either side and has no strongly connected users, but has a strong peak in activity during one particular hour in the evening of $27^{th}$ of November.
This peak in activity is not present in other communities, and does not follow immediately after publication of any online news articles, which makes it highly likely that it originated because of the peer-driven influence exclusively.
This is further reinforced by the fact that majority of users in this community come from a small town in Croatia as well as the major university towns in Croatia, which makes it highly likely that these are indeed users who form a strong real-life community of friends.
We will consider this peak in activity as a gold standard for peer-driven influence.

\subsection{Mass media -- external influence}
\begin{figure*}
\centerline{\includegraphics[width=\textwidth]{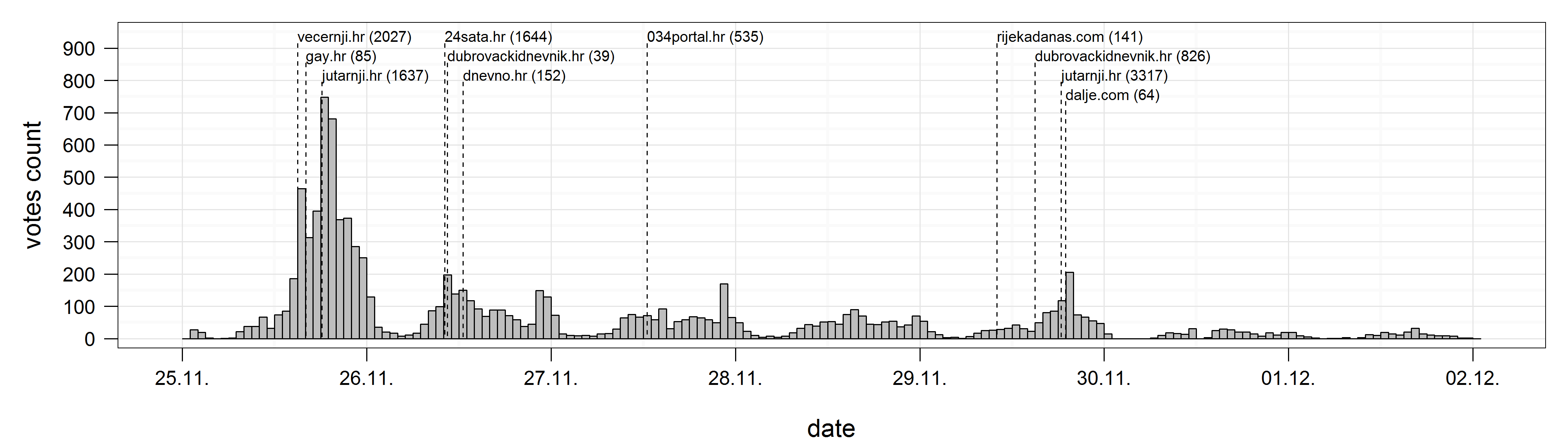}}
\caption{
Hourly count of votes.
Voting application opened for public at $25^{th}$ of November at $1:24$ AM. 
We stopped collecting votes at midnight $2^{nd}$ of December. 
Plots are annotated with times and domains of online news articles that reported on our application, along with the number of visitors that visited our web page through referral from these domains through the whole period. 
Many large peaks in votes numbers correspond closely to the publication times of major news articles. 
Peaks that do not have such correspondence are probably due to the dynamic of social referrals. 
One way to demonstrate this is to show that majority of votes from one particular peak came from a particular community of highly connected friends. We show later that this is the case with the peak at around 11 PM, $27^{th}$ of November.}
\label{fig:hourly_votes_annotated}
\end{figure*}

As a proxy of external influence we use online news articles that reported on our application, and we weight them by the number of visitors that visited our application through referral from these domains through the whole period. 
We retrieved information on online news articles during one week prior to the referendum by observing referral traffic to our site and the total number of visitors obtained by the Google Analytics.
Total number of visitors gives us an rough estimate on the \emph{external} influence each particular news article had in motivating the users for voting.
From Figure~\ref{fig:hourly_votes_annotated} it is immediately obvious that majority of online news articles are followed with large peaks in voting activity.
This reinforces our hypothesis that media had a large influence in activating the users.
We believe that this \emph{external} influence has a distinct pattern that can be distinguished from the \emph{peer} influence mediated within the network of friendships.
We will use the exact times of news articles as a gold standard when evaluating the external influence.

\section{Modeling peer and external influence}
\label{sec:model}

In this manuscript, we are modeling the activation of users in the online social network. 
In our case the activation of a user represents expressing the opinion as a form of social engagement on our web-site prior to the December 1st 2013 referendum in Croatia.
User activations are moderated by the superposition of peer influences in social network and the external influence from mass media. 
We assume that each activated node $i$ transfers the peer activation influence $p_0$ to its neighbors, which decays exponentially in time: $p_0 e^{-\lambda t}$ with the decay parameter $\lambda$. 
Each activated node can independently transfer the influence to non-activated node. 
Then for each node $i$ at time $t$, the probability of the activation from its already activated neighbors $N(i)$ is:  
\begin{equation}
p_i(t) = 1 - \prod_{k \in N(i):t_k<t} (1-p_{0}e^{-\lambda (t-t_k)}),
\label{eq:probability_of_peer_activation}
\end{equation}
where $t_k$ denotes the time of activation of neighboring node $k$ which activated before time $t$.  
%At each time $t$, the set of nodes in the social network $G=(V,E)$ is partitioned to the set of already activated nodes $A_t = \left\{ v_i \in V: t_i < t \right\}$ and set of non-activated nodes $A_t^c = \left\{ v_i \in V: t_i > t \right\}$.
%In the time window $[t-\Delta,t]$ some subset of non-activated nodes become active and we denoted them as the set difference $W_{t+\Delta} = A_{t+\Delta} \setminus A_t$. 

\begin{figure}
\centerline{\includegraphics[width=0.5\textwidth]{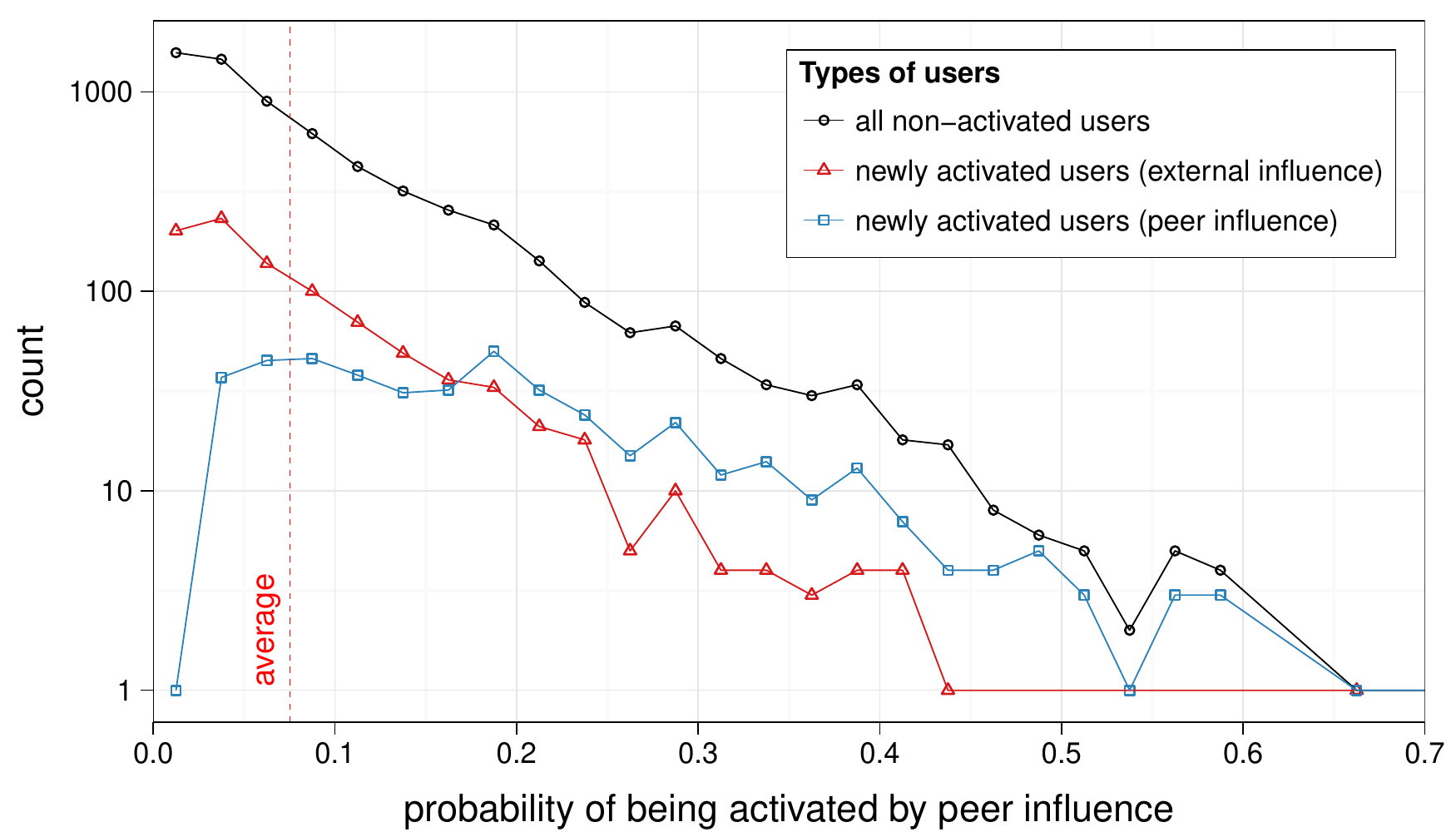}}
\caption{
Distribution of probability of being activated by any of your peers ($p_i(t)$ from equation~\ref{eq:probability_of_peer_activation}) for all non-activated users and newly activated users that activated through either peer or external influence.
This distribution is taken from the $15^{th}$ step of the simulation described in section~\ref{sec:simulations}.
Proportion of externally-activated users is constant regardless of the user's probability of being activated by peer influence.
On the other hand, proportion of peer-activated users rises proportionally with their probability of being activated by peer influence.
Vertical dashed line show the average probability of being activated by any of your peers $p_i(t)$ for all non-activated users ($\mu(t)$ from equation~\ref{eq:peer_influence}).
Out of all newly activated users whose $p_i(t)$ is \emph{lower} than the $\mu(t)$, majority of them are activated due to the external influence.
In contrast, out of all newly activated users whose $p_i(t)$ is \emph{higher} than the $\mu(t)$, majority of them are activated due to the peer influence.
This justifies our approach for distinguishing peer and external influence described in equation~\ref{eq:peer_influence}.
}
\label{fig:influence_distribution}
\end{figure}

Next, we calculate the expected probability of activation over all non-activated users at time $t$ and denote it with $\mu(t)$. 
\begin{equation}
\mu(t) = \frac{1}{ N }\sum_{i: t_i \in (t, +\infty)} p_i(t) ,
\label{eq:average_influence}
\end{equation}
where $N$ denotes the number of non-activated users at time $t$. 
The external influence is estimated in a non-parametric way, as every activation which can not be explained with the peer activation.
Next, we assume that the external influence is distributed more uniformly around the network than the peer influence, simply because the mass media can influence very large number of individuals at the same time.
Nodes that activated recently, in time window $[t-\Delta,t]$, we call the newly activated nodes.
If there is only a external influence present with uniform influence, the set of newly activated nodes should resemble to the unbiased uniform sub-sample of the set of all non-activated nodes. 
But, if there exists a significant peer influence, the set of newly activated nodes should be a biased sub-sample over the set of all non-activated nodes as the peer influence is network localized.
Formally, we estimate this peer bias in time segment $[t-\Delta,t]$ as a sum of all the activation probabilities for all the newly activated users subtracted by the expected probability of activation for the non-activated users.
\begin{equation}
peer(t) = \sum_{i: t_i \in [t-\Delta,t]} \mathbbm{1}(p_i(t) - \mu(t)),
\label{eq:peer_influence}
\end{equation}
where the $\mathbbm{1}(x)$ denotes indicator function which is equal to $1$ if the argument is non-negative, otherwise it is zero.
If the newly activated node $i$ has probability lower than the $\mu(t)$ we classify it as an external activation node.  
Figure~\ref{fig:influence_distribution} (obtained by the simulations) shows that users who activated due to the external influence have $p_i(t)$ distributed as an uniform unbiased sub-sample of the set of all non-activated nodes probabilities.
%Figure~\ref{fig:influence_distribution} shows that users who activated due to the external influence tend to have $p_i(t)$ that is \emph{lower} than the $\mu(t)$, while users who activated due to the peer influence tend to have $p_i(t)$ is \emph{higher} than the $\mu(t)$, which justifies our approach for distinguishing peer and external influence described in equation~\ref{eq:peer_influence}.
As a baseline for external influence we use a method~\cite{Myers2012,Gomez-Rodriguez2011} that classifies an activation as external if the user had no previously activated friends.
This is a conservative measure that tends to underestimate the true external influence~\cite{Myers2012} because after majority of users is activated it is extremely unlikely for newly activated users to have no previously activated friends, even if they really are activated by an external influence.
In our case the quantity $\mu(t)$ increases as time progresses and thus overcomes this limitation of underestimating the external influence of the baseline solution.
Note that when the $\lambda = 0$ we obtain the baseline model for external influence.

\section{Experiments}

We propose a new method that estimates peer and external influence using information on friendship network and activation cascade.
We evaluate our method on both simulated dynamics and actual dynamics using before mentioned gold standards.

\subsection{Evaluation on simulated dynamics}
\label{sec:simulations}
\begin{figure}
\centerline{\includegraphics[width=0.5\textwidth]{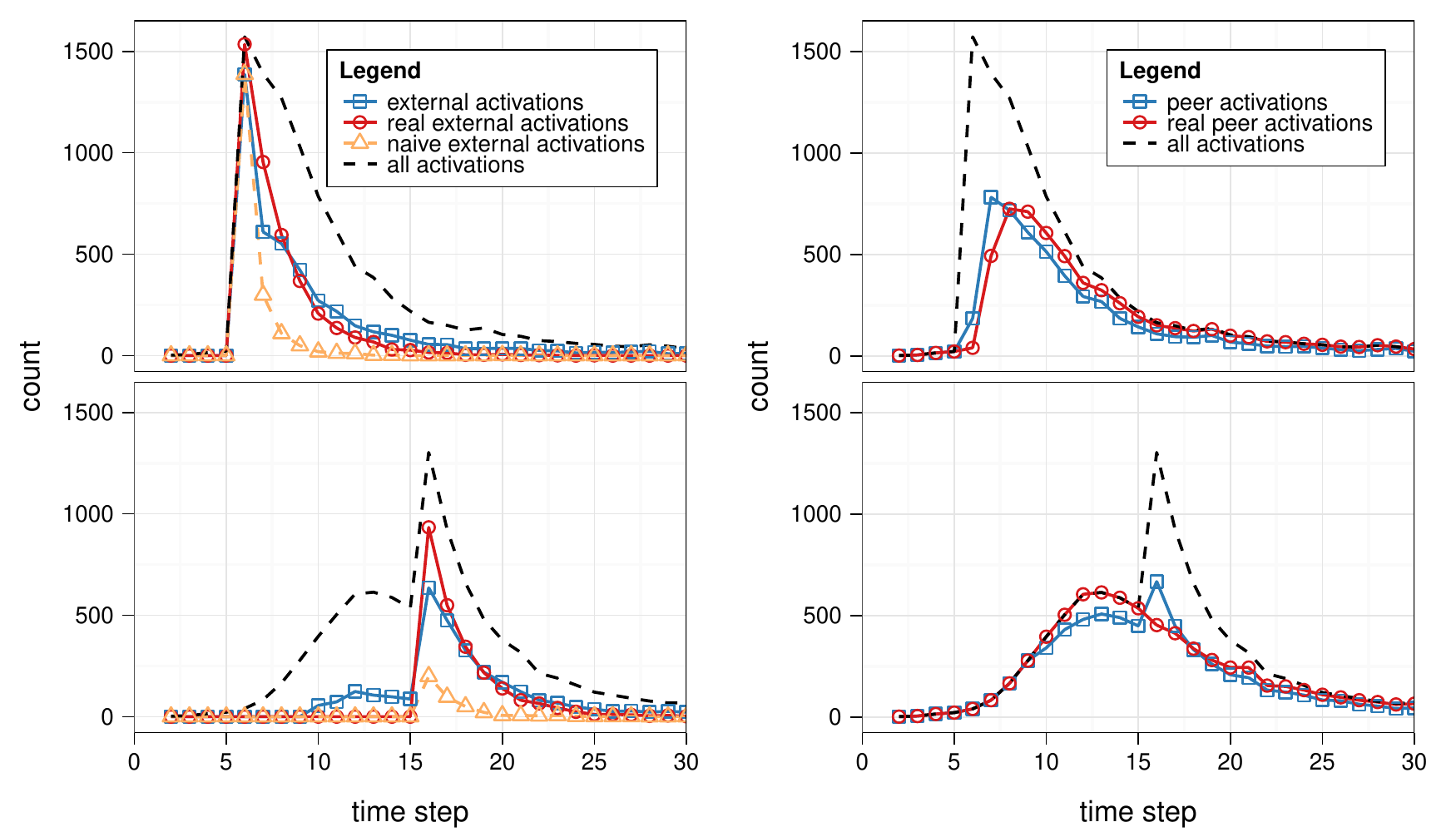}}
\caption{Simulated voting dynamics on referendum network. 
We simulated voting dynamics on referendum network using epidemic model with exponential decay for both peer influence and external influence.
Our method is able to estimate magnitudes of external influence (left panels) and peer influence (right panels) in cases where external influence is firing at the $5^{th}$ (top panels) and $15^{th}$ (bottom panels) step.
}
\label{fig:simulated_dynamics}
\end{figure}

For simulating voting dynamics we use an actual friendship network and a discrete epidemic model where each node upon activation starts spreading exponentially decaying influence of the form $p_{0} e^{-\lambda_{p}(t-t')}$ to all its neighbors.
This influence translates directly to the probability of activating any yet non-activated friend in the next time step.
We initialize simulation with a single activated user.
The simulation then progresses in discrete steps with every user having an independent probability to be activated by any of its already activated friends in each step.
We also introduce external influence that determines the probability of activation uniformly for all yet non-activated users in the next time step, regardless of how many of their friends are already activated.
In simulation mode, we use the functional form for the external influence, which is similar to the peer influence - with spiked exponentially decaying influence $q_{0} e^{-\lambda_{e}(t-t')}$.
%Note, that we estimate the external influence in a non-parametric way, but in simulation mode we set the functional form of external influence.
Figure~\ref{fig:simulated_dynamics} shows results on simulated voting dynamics on referendum network with peer influence parameters: $p_{0} = 0.03$, $\lambda_{p} = 0.02$ and external influence parameters: $q_0 = 0.2$, $\lambda_{e} = 0.3$ that fires at $5^{th}$ and $15^{th}$ step of the simulation.
Using our method outlined in section~\ref{sec:model} we are able to estimate the total number of users activated due to the peer or the external influence.
In comparison, baseline for external influence tends to underestimate total number of externally activated users, especially after some part of the network is already activated. 

\subsection{Evaluation on actual activation cascade}
\begin{figure*}
\centerline{\includegraphics[width=\textwidth]{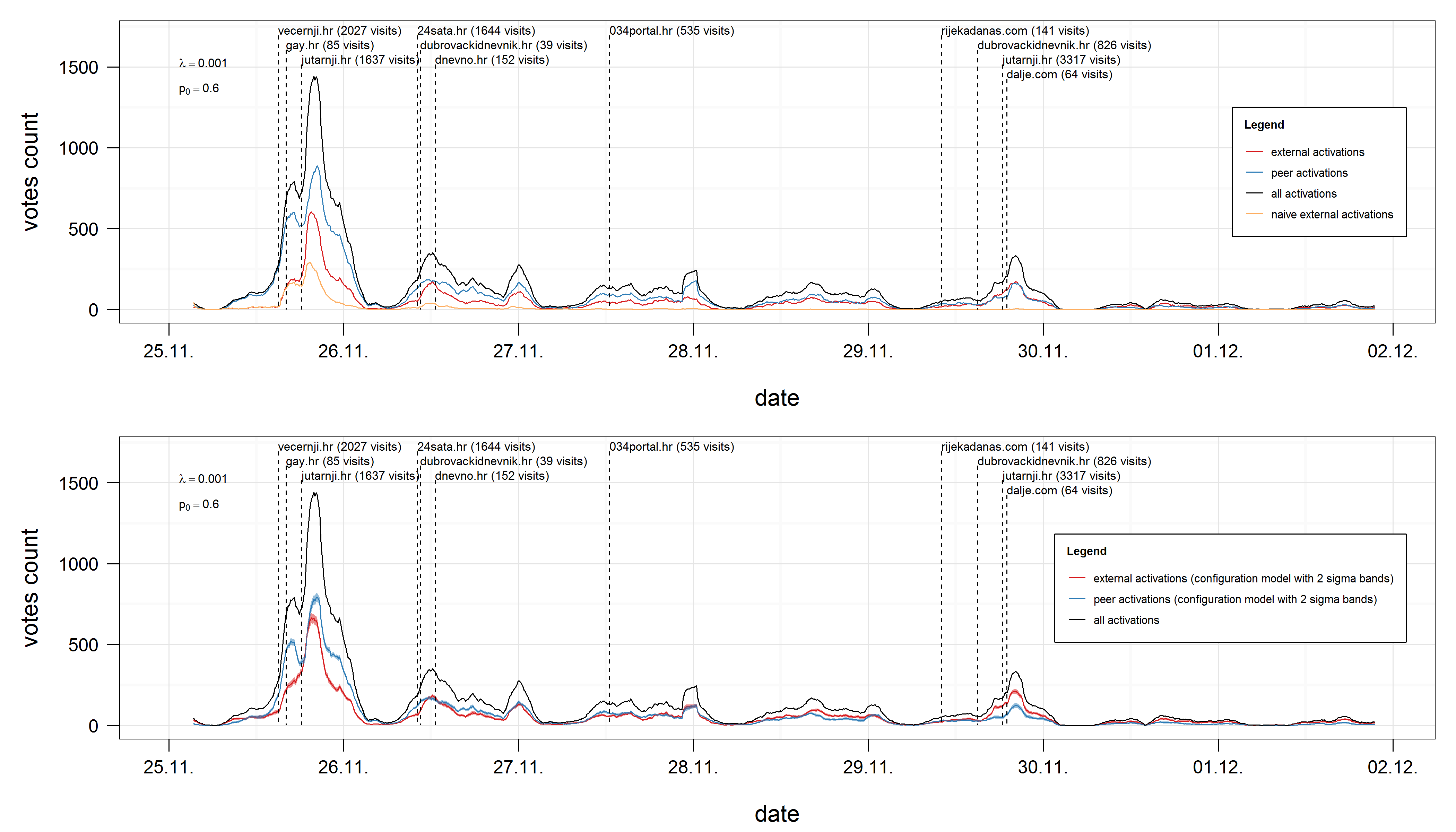}}
\caption{Evaluating external influence detection on real referendum activation cascade. 
We characterize activation of each node as \emph{peer} or \emph{external} using assumption of exponentially decaying peer influence each user has at the time of activation.
Top panel shows the total number of peer and external activations as estimated by our method.
Bottom panel shows evaluation of our method on the configuration model of the network with the actual activation cascade.
As expected, in this case the effect of peer influence is reduced, allowing the external influence to dominate even during highly peer-driven periods like the evening of $27^{th}$ of November.
In our evaluation we choose the parameters of peer influence $p_{0} = 0.6$ and $\lambda_{p} = 0.001$, and we used sliding window of two hours to define newly activated users.}
\label{fig:referendum_activations_epidemic}
\end{figure*}

We finally use the method we outlined in section~\ref{sec:model} to estimate the magnitude of peer and external influence using the actual Facebook friendship network and voting times of users.
Figure~\ref{fig:referendum_activations_epidemic} shows the estimated number of peer and external activations aggregated in two hour sliding window throughout the voting period.

\begin{figure}
\centerline{\includegraphics[width=0.5\textwidth]{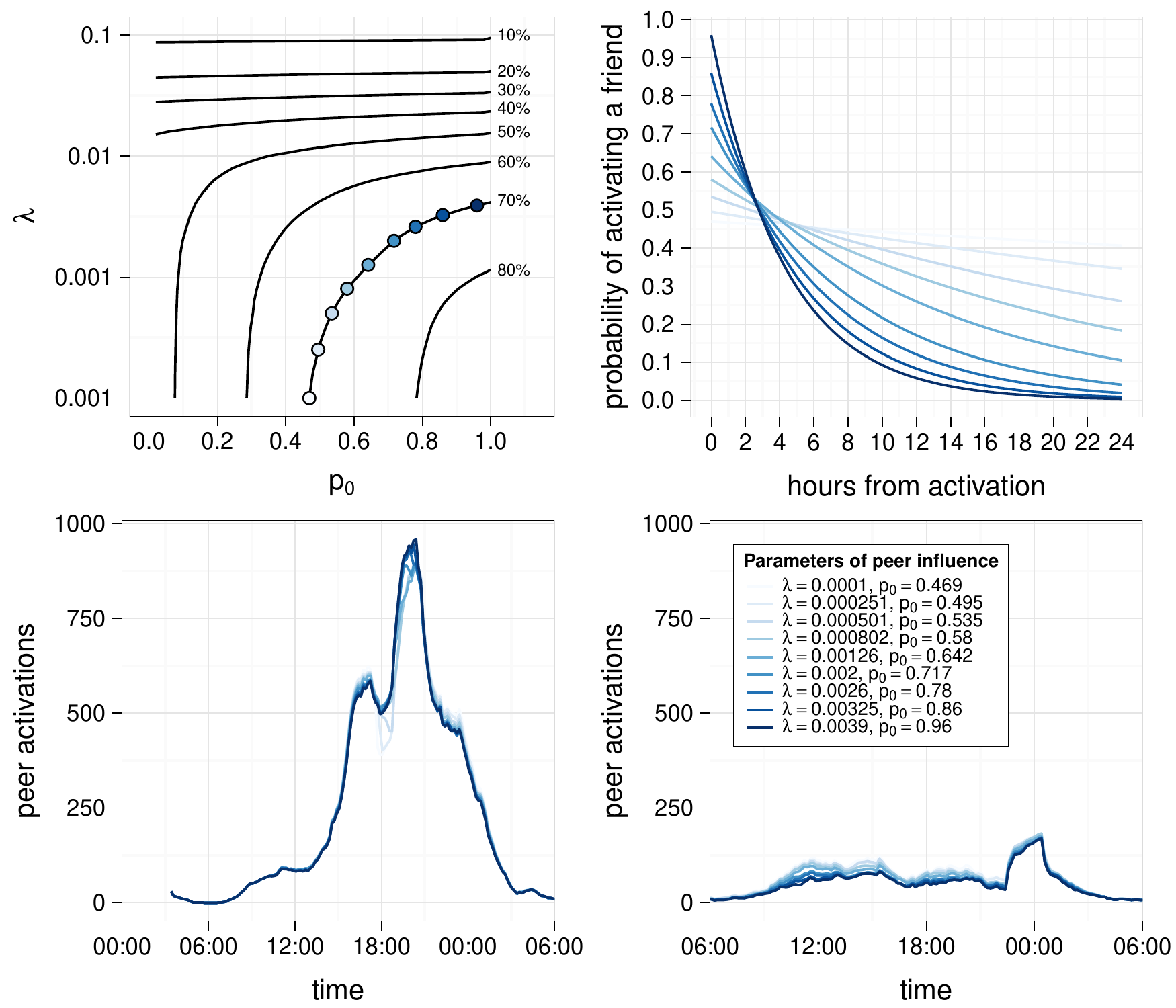}}
\caption{
Choosing optimal parameters $\lambda$ and $p_{0}$ that determine exponential decay of peer influence $p_{0} exp(-\lambda(t-t'))$.
Lower values of $\lambda$ and higher values of $p_{0}$ correspond to the strong and slow decaying peer influence, raising the fraction of peer-activated users.
Higher values of $\lambda$ and lower values of $p_{0}$ correspond to the weak and faster decaying peer influence, lowering the fraction of peer-activated users.
We choose the values so that the total fraction of peer-activated users in the first day of voting corresponds to the fraction of users visiting our web application via links from Facebook, which is in our case around 70\%.
Top left panel shows contour plot of the parameter space, with the percentage of peer-activated users during the first day on the right.
All $(\lambda,p_{0})$ pairs lying on the curve of $70\%$ are optimal according to the criterion given above.
Their corresponding peer influence curves are plotted on the top left panel.
Bottom panels show that these parameters all produce similar peer activation curves, as can be seen on the bottom left panel for Monday and on the bottom right panel for Wednesday.
}
\label{fig:optimizing_lambda}
\end{figure}

\textbf{Optimizing parameters of peer influence.}
In order to choose appropriate parameters of peer influence $\lambda$ and $p_{0}$ we exploit the information on the visitors to our web site we have from Google Analytics.
Out of total of $25154$ visitors that visited our web site during the first day of voting, $17587$ came by referral from Facebook, while the rest came by referrals from various news articles external to Facebook. 
This gives us a rough estimate of the ratio of peer and external visitors to our web site.
We choose parameters $\lambda$ and $p_{0}$ so that the ratio of peer-activated users during the first day of voting approaches $70\%$, as shown on Figure~\ref{fig:optimizing_lambda}.
There are many possible pairs $(\lambda,p_{0})$ that are optimal according to the criterion given above, so we choose $p_{0} = 0.6$ and $\lambda_{p} = 0.001$ as an illustrative example on the Figure~\ref{fig:referendum_activations_epidemic}. But, note that our estimation method of peer and external influence is quite robust to range of $(\lambda,p_{0})$ parameters (see Figure~\ref{fig:optimizing_lambda}, bottom panels).

\textbf{Validating the estimate of peer and external influence.}
In order to validate estimated magnitudes of peer and external influence we use couple of gold standards.
As a gold standard for external influence we use publication times of online news articles that mention our application.
Number of visitors that visited our web site by referrals from these domains gives us an estimate of the relative magnitude of influence of each external event.
We observe a noticeable rise in external influence immediately after publication of each online news articles, and its decay after time.
On the other hand, as a gold standard for peer influence we use: (i) initial dynamics that occurred before the first online news article, (ii) time-localized dynamics that originates from a single well defined community on the night of $27^{th}$ of November.
As shown on the top panel of Figure~\ref{fig:referendum_activations_epidemic}, the magnitude of external influence before the first online news article is negligible, with sharp rise just after the publication of the first online news article.
The peer influence remains dominant throughout the first day.
In comparison, baseline method correctly estimates the magnitude of the first peak of external activations, but it quickly starts to underestimate it, and it fails to identify external activations completely after the first day.
This is due to its overconfident assumption that newly activated users will have no activated friends if they are activated due to the external influence, which is hard to satisfy as soon as the finite size network becomes saturated with activated nodes.
Another period where we expect high peer influence is during the night of $27^{th}$ of November.
As we showed in section~\ref{sec:communities} and Figure~\ref{fig:community}, this period exhibits unusually large voting activity originating from a single well defined community of users.
Indeed, there is a sharp rise in peer influence during few hours of the evening, while external influence remains flat.

\textbf{Configuration model of friendship network.} 
We evaluate our method on configuration model of the network while keeping the actual activation times.
Configuration model produces an ensemble of networks by rewiring all friendship connections so that each user preserves its total number of friendships.
This preserves the global topological properties of the network like degree distribution, but disrupts mesoscale and local properties like communities and individual friendships that mediate peer influence.
In this case we expect the majority of peer influence to be wrongly misinterpreted as external influence, as really is the case on the bottom panel of Figure~\ref{fig:referendum_activations_epidemic}.
Rewiring friendship connections in the configuration model decouples the activation cascade from the actual network, and majority of peer influence will spread out across the network and be interpreted as external influence.
Two specific cases illustrate this clearly.
First, before the publication of the first online news article majority of activations are due to the peer influence, but in configuration model the peer influence is of equal magnitude as the external influence in this period.
Similar, on the night of $27^{th}$ of November we know that majority of activations came from a single well defined community of users, meaning that peer influence should dominate, but in configuration model we again have equal magnitudes of peer and external influence.

\section{Discussion}

Our analysis show that, under the assumption on exponentially decaying peer influence, it is possible to estimate magnitude of external and peer influence in social networks using information on friendship network and the times of activation.
This is possible due to the different mechanics of how external and peer influence propagate through network - external influence is a sort of mean-field effect that targets all users uniformly, while peer influence propagates from recently activated users.
By exploiting this property we were able to give a reasonable estimate of the magnitude of peer and external influence in the Facebook network of users who registered on our application and voted on the upcoming referendum question.
Of course, there are many uncertainties in the data we collected, especially regarding the motivations of our users and the information diffusion pathway between them.
Friendship network is not complete as we have only friendship relationships between users who registered on our application, and we do not have data on many other plausible pathways of peer influence like word-of-mouth, email and other social networks.
Our analysis would certainly benefit from a more detailed data on Facebook and web browsing activity of users of our application.

\section{Acknowledgments}

The work is supported in part by Croatian Science Foundation (grant no. I-1701-2014) and by the EU-FET project MULTIPLEX (grant no. 317532).
We would like to thank the people who actively collaborated in the development of the Facebook application for the collection of data: Bruno Rahle, Tomislav Lipi\'{c}, Vedran Ivanac and Matej Mihel\v{c}i\'{c}.
Also, many other people with whom we had fruitful discussions: Vinko Zlati\'{c}, Sebastian Krausse. 

% \bibliographystyle{abbrv}
\bibliographystyle{unsrt}
% \bibliography{referendum}  % sigproc.bib is the name of the Bibliography in this case

% ACM needs 'a single self-contained file'!
%

\end{document}